\documentclass{moriond}
\usepackage[applemac]{inputenc}
\usepackage{amssymb}
\def\Journal#1#2#3#4{{#1} {\bf #2}, #3 (#4)}

\def\PLB{{\em Phys. Lett.}  B}

\newcommand{\lsim}{\raisebox{-0.13cm}{~\shortstack{$<$ \\[-0.07cm] $\sim$}}~} 
\newcommand{\gsim}{\raisebox{-0.13cm}{~\shortstack{$>$ \\[-0.07cm] $\sim$}}~} 
\newcommand{\beq}{\begin{eqnarray}} 
\newcommand{\eeq}{\end{eqnarray}} 
\newcommand{\tb}{\tan \beta}

\def\be{\begin{equation}}
\def\ee{\end{equation}}
\def\bea{\begin{eqnarray}}
\def\eea{\end{eqnarray}}

\newcommand{\Photo}{}

\begin{document}
\begin{flushright}
LPT-Orsay-14-24
\end{flushright}

\vspace*{4cm}
\title{SIMPLIFIED DESCRIPTION OF THE MSSM HIGGS SECTOR}

\author{ J\'er\'emie Quevillon }

\address{Laboratoire de Physique Théorique d'Orsay, Bâtiment 210, Université Paris Sud 11,\\ 91405 Orsay Cedex, France}

\maketitle\abstracts{
In the Minimal Supersymmetric extension of the Standard Model or MSSM, the lighter Higgs
boson has a rather large mass, $M_h\approx 125$ GeV. Together with the non-observation of superpartners at the LHC, this suggests that the  SUSY--breaking scale is rather high, 
${M_{S} \gsim 1}$~TeV. This implies a dramatic simplification of the MSSM Higgs sector 
that is summarised here.   
}

\section{The post-Higgs boson discovery MSSM Higgs sector}

In the MSSM, two Higgs doublets $H_{d}$ and $H_{u}$ are needed to break the electroweak symmetry, leading to three neutral and two charged Higgs states;  for a review see 
Ref.\cite{Djouadi:2005gj}. The tree--level masses of the CP--even $h$ and $H$  
bosons depend only on $\tb = v_d/v_u$, the ratio of vevs of the two doublets and on 
the pseudoscalar Higgs mass $M_{A}$. Nevertheless, many parameters of the MSSM such 
as the SUSY scale,  taken to be the geometric average  of the stop masses $M_S\!= 
\!\sqrt {m_{\tilde t_{1}} m_{\tilde t_{2}} }$, the higgsino mass  $\mu$ and the 
stop/bottom trilinear couplings $A_{t/b}$ enter $M_{h/H}$ through loop corrections.  
The CP--even Higgs  mass matrix  can be written in the basis as:
\beq
{\cal M}_{S}^2=M_{Z}^2
\left(
\begin{array}{cc}
  c^2_\beta & -s_\beta c_\beta \\
 -s_\beta c_\beta & s^2_\beta \\
\end{array}
\right)
+M_{A}^2
\left(
\begin{array}{cc}
 s^2_\beta & -s_\beta c_\beta \\
 -s_\beta c_\beta& c^2_\beta \\
\end{array}
\right)
+
\left(
\begin{array}{cc}
 \Delta {\cal M}_{11}^2 &  \Delta {\cal M}_{12}^2 \\
 \Delta {\cal M}_{12}^2 &\Delta {\cal M}_{22}^2 \\
\end{array}
\right)
\label{eq1}
\eeq
where we use the notation $c_\beta \! \equiv \! \cos\beta$, $s_\beta \! 
\equiv \! \sin\beta$ and include the radiative corrections into a $2\times 2$ matrix  
$\Delta {\cal M}_{ij}^2$. One can then easily derive the Higgs  
masses $M_{h,H}$ and the mixing angle $\alpha$ that diagonalizes the $h,H$ 
system, $h= -\sin\alpha H_d^0 + \cos\alpha H_u^0$ and $H= \cos\alpha H_d^0 + \sin\alpha H_u^0$:
\begin{eqnarray}
\hspace{-1.0cm}
M_{h/H}^2&=&\frac{1}{2} \big( M_{A}^2+M_{Z}^2+ \Delta {\cal M}_{11}^2+ 
\Delta {\cal M}_{22}^2  \mp  \sqrt{ M_{A}^4+M_{Z}^4-2 M_{A}^2 M_{Z}^2 
c_{4\beta} +C} \big) \\
\hspace{-1.0cm}
\tan \alpha&=&\frac{2\Delta {\cal M}_{12}^2 - (M_{A}^2 + M_{Z}^2) s_{\beta}}
{ \Delta {\cal M}_{11}^2 -  \Delta {\cal M}_{22}^2 + (M_{Z}^2-M_{A}^2)
c_{2\beta} + 
\sqrt{M_{A}^4 + M_{Z}^4 - 2 M_{A}^2 M_{Z}^2 c_{4\beta} + C}}
\end{eqnarray}
\vspace*{-5mm}
\begin{eqnarray}
C=  4 \Delta {\cal M}_{12}^4\! + \!( \Delta {\cal M}_{11}^2 \!- \! 
\Delta {\cal M}_{22}^2)^2 \!- \! 
 2 (M_{A}^2 \! - \! M_{Z}^2)( \Delta {\cal M}_{11}^2 \! - \! \Delta M_{22}^2) 
 c_{2\beta} \!   - \!
 4 (M_{A}^2 \! + \!M_{Z}^2)  \Delta {\cal M}_{12}^2 s_{2\beta} \nonumber
\end{eqnarray}

In previous works~\cite{Maiani,O1}, it was pointed out that since the measured 
value of the $h$ boson mass is high, $M_{h}=125$~GeV, leading to a rather large 
SUSY-breaking scale~\cite{Arbey:2011ab}, $M_S \gsim 1$ TeV, it implies that the 
leading radiative corrections are now almost fixed when the constraint $M_{h}=125$~GeV 
is taken into account. In the $2\times 2$ correction matrix of eq.~(\ref{eq1}), 
only the $\Delta{\cal M}^{2}_{22}$ entry 
which involves the by far leading top/stop corrections proportional to 
the fourth power of the top Yukawa coupling, is relevant to a good 
approximation~\cite{Djouadi:2013uqa}.  
In this limit $\Delta{\cal M}^{2}_{22} \gg \Delta{\cal M}^{2}_{11}, \Delta{\cal M}^{2}_{12}$, one can simply trade 
$\Delta {\cal M}^{2}_{22}$ for the known $M_h$ value:   
\beq
\Delta {\cal M}^{2}_{22}= \frac{M_{h}^2(M_{A}^2  + M_{Z}^2 -M_{h}^2) - M_{A}^2 M_{Z}^2 c^{2}_{2\beta} } { M_{Z}^2 c^{2}_{\beta}  +M_{A}^2 s^{2}_{\beta} -M_{h}^2}\, .
%or \Delta {\cal M}^{2}_{22}= \frac{-M_{A}^2 M_{h}^2 + M_{h}^4 - M_{h}^2 M_{Z}^2 + M_{A}^2 M_{Z}^2 \cos^{2}2 \beta} {M_{h}^2 - M_{Z}^2 \cos^{2}\beta - M_{A}^2 \sin^{2}\beta}
\eeq
In this case, called habemus MSSM or hMSSM in Ref.\cite{Djouadi:2013uqa},  one obtains 
simple expressions for the mass $M_H$ and the angle $\alpha$ in terms of 
$M_A,\tb$ and $M_{h}$:
\beq
{\rm hMSSM}:~~ 
\begin{array}{l} 
M_{H}^2 = \frac{(M_{A}^2+M_{Z}^2-M_{h}^2)(M_{Z}^2 c^{2}_{\beta}+M_{A}^2
s^{2}_{\beta}) - M_{A}^2 M_{Z}^2 c^{2}_{2\beta} } {M_{Z}^2 c^{2}_{\beta}+M_{A}^2
s^{2}_{\beta} - M_{h}^2} \\
\ \ \  \alpha = -\arctan\left(\frac{ (M_{Z}^2+M_{A}^2) c_{\beta} s_{\beta}} {M_{Z}^2
c^{2}_{\beta}+M_{A}^2 s^{2}_{\beta} - M_{h}^2}\right)\, .
\end{array}
\label{wide} 
\eeq
Concerning the charged Higgs boson, the quantum corrections to its mass are much smaller for large $M_{A}$, and one can write to a good approximation, $M^2_{H^\pm} \simeq 
M_A^2 + M_W^2$.

This approach allows to disregard the radiative corrections in the MSSM Higgs sector and their complicated dependence  on all the MSSM parameters.
This considerably simplifies the phenomenological studies in the MSSM Higgs sector which up to now do not use the constraint $M_{h}=125$ GeV as an input as it should be, and rely either on benchmark scenarios in which most of the MSSM parameters are fixed or refuge to large scans over the parameter space.

\section{Fit of the SM Higgs couplings}

In the MSSM, the couplings of the lighter $h$ state 
to gauge bosons and fermions, normalized to their SM values read:
\beq  
c_V^0  \! = \!  \sin(\beta \!-\! \alpha)  \ ,  \ \  c_t^0 \!  = \!  
\frac{\cos\alpha}{\sin\beta} \ ,  \ \  c_b^0 \!  = \!  -
\frac{\sin\alpha}{\cos\beta} \,.  \label{eq6} 
\eeq
They depend on the tree--level inputs $\tb$ and $M_{A}$ but also on the full 
MSSM spectrum because of the quantum corrections that enter the angle $\alpha$ as in 
the case of the Higgs masses. As discussed earlier, knowing $\tb$ and $M_{A}$ and 
fixing $M_h$ to its  measured 
value, the couplings can be determined. Nevertheless, this applies only for the radiative corrections to the 
Higgs masses. In addition, there exists direct radiative corrections to the Higgs couplings different from the ones 
of the mass matrix in eq.~(\ref{eq1}) and which will complicate the situation. 

\begin{figure}[!h]
\begin{center}
%\vspace*{-6.9cm}
%\hspace*{-2cm}
\begin{tabular}{cc}
\includegraphics[scale=0.48]{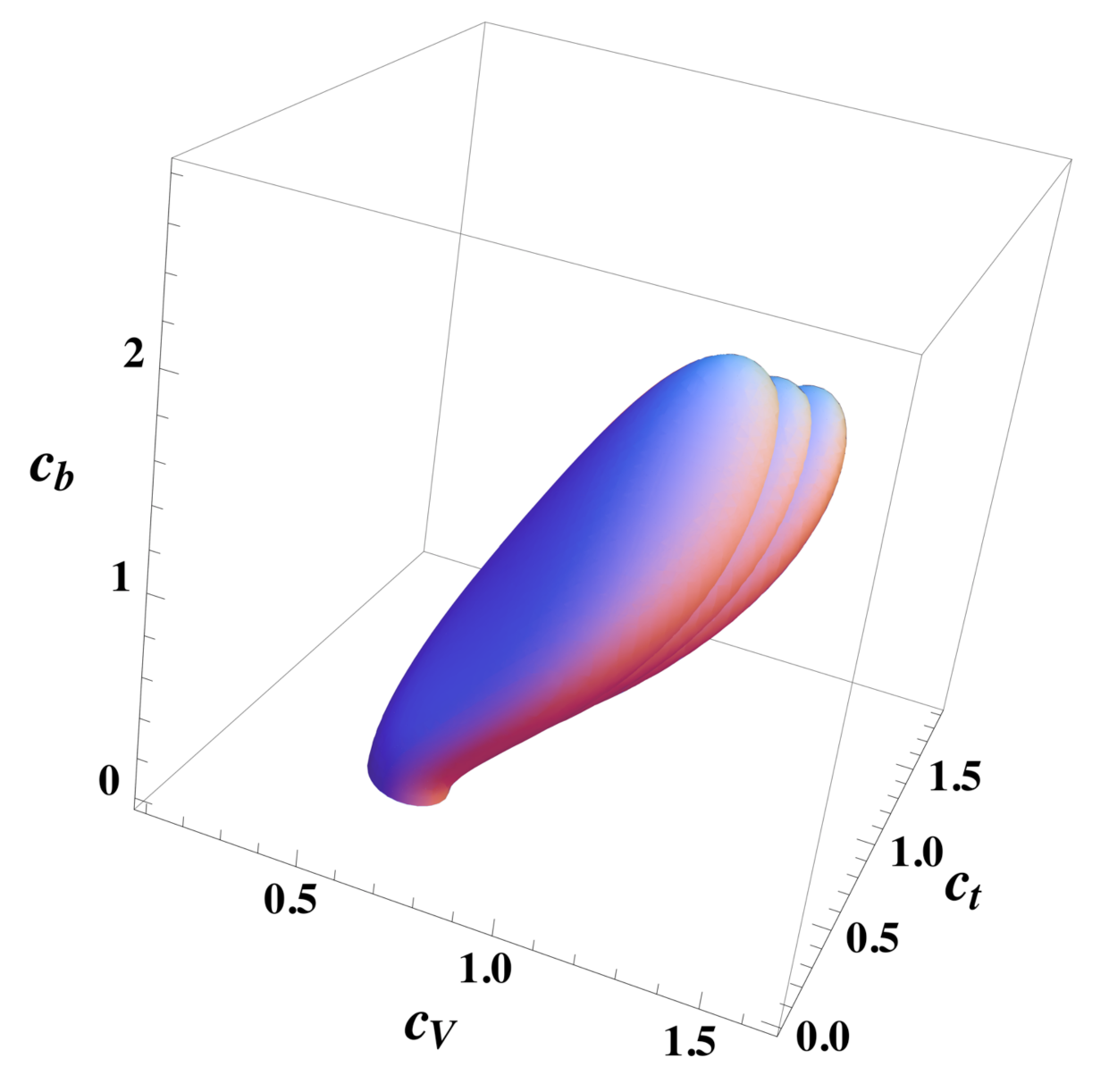} & \hspace*{1cm}
\includegraphics[width=7.0cm,height=5.5cm]{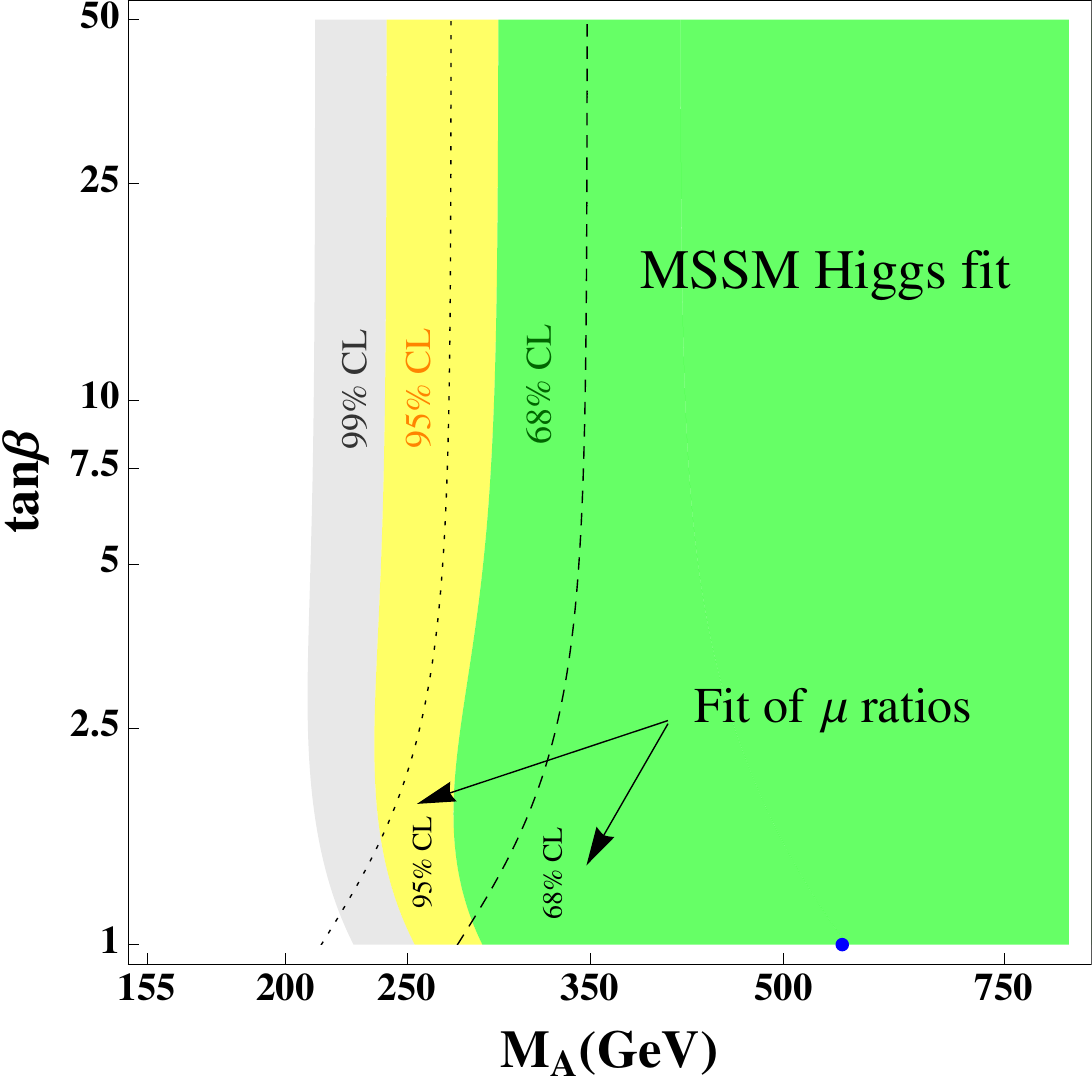}
\end{tabular}
\end{center}
%\vspace*{-6.7cm}
\caption[]{Left: 
best-fit regions at 99\%CL for the Higgs
signal strengths in the three dimensional space $[c_t, c_b, c_V]$~\cite{Djouadi:2013uqa}. 
Right: best-fit regions for the signal strengths and their ratios in the 
plane $[\tan \beta, M_A]$; the best point is in  blue~\cite{Djouadi:2013uqa}.}
\label{fig1}
%\vspace*{-.6cm}
\end{figure}

If the $h$ coupling to the bottom and top quarks could be significantly modified
(by stop loops in the production process $gg\to h$ in the former  and by 
the $\Delta_b$ corrections in the latter cases; see Ref.\cite{Djouadi:2013uqa}), 
$c_{t,b}^0 \to c_{t,b}$, the 
couplings to $\tau$ leptons and $c$ quarks do not receive substantial direct 
corrections and one still has $c_{c,\tau} \approx c_{t,b}^0$.
Consequently, because of the direct radiative corrections,  the Higgs couplings cannot be described 
by only $\beta$ and $\alpha$ as in eq.~(\ref{eq6}). To characterize 
the Higgs particle at the LHC, it was advocated~\cite{Djouadi:2013uqa} that three independent 
$h$ couplings should be considered, namely $c_{t}$, $c_{b}$ and $c_{V}=c_V^0$. 
Thus, one can define the following effective Lagrangian:     
\beq
{\cal L}_h  &\! =\! &   c_V  g_{hVV}  h  V_{\mu}^+ V^{- \mu} 
+% +  c_V  g_{hZZ}  h Z_{\mu}^0 Z^{0 \mu} 
\label{Eq:LagEff} %\\ &\!-\! &   
  c_t y_t  h  \bar t_L  t_R \! -\!  c_t  y_c h  \bar c_L  c_R \! -\! 
  c_b  y_b   h  \bar b_L b_R  \! - \! c_b y_\tau h  \bar \tau_L \tau_R 
 \! + \! {\rm h.c.}  
\eeq
where $y_{t,c,b,\tau}=m_{t,c,b,\tau}/v$ are the Yukawa couplings of the heavy
SM fermions, $g_{hVV}\! = \! 2M^2_V/v$ the $hVV$ couplings with $V\!=\!W,Z$. 
Following an earlier analysis performed in Ref.\cite{Fits} where details can be found, 
a three--dimensional fit of the $\sqrt s=7+$8 TeV ATLAS and CMS Higgs data has been
performed and the result in the space $[c_t, c_b, c_V]$ is shown 
on the left-hand side  of Fig.~\ref{fig1}. The obtained best-fit values for the 
Higgs couplings are: $c_t=0.89$,~$c_b=1.0$ and $c_V=1.02$.

In cases where the direct corrections are not quantitatively significant 
one can reduce the number of effective parameters down to two using the MSSM relations of 
eq.~(\ref{eq6}). 
Using the formulae of eq.~(\ref{wide}) for the mixing angle and the $M_h \! \approx 125$ GeV value as an input, 
one can perform a fit in the $[\tb,  M_{A}]$ plane as shown on the right-hand side of Fig.~\ref{fig1}. It illustrates the 
68\%, 95\% and  99\%CL contours obtained from fitting the signal strengths and their ratios. 
The best-fit point is realized for the values $\tan\beta\!=\! 1$  and $M_A \! = \! 557 \; {\rm GeV}$, 
which translates into $M_{H}= 580$~GeV, $M_{H^\pm}= 563$~GeV and $ \alpha=-0.837~{\rm rad}$. Such a 
low $\tb$ point implies an extremely large SUSY scale value, $M_{S} = {\cal O}(100)$ 
TeV to accommodate a $125$ GeV Higgs boson. Notice, that the $\chi^2$ value is relatively flat all over the 
$1\sigma$ region and, thus, larger $\tb$ values could also
be appropriate, hence allowing for not too large SUSY scale values. Nevertheless, 
one obtains that the pseudoscalar should verify $M_{A} \gsim 200$ GeV in all cases.

\section{Heavy scalar searches}

\begin{figure}[!h]
\begin{center}
%\vspace{-7.cm}
\begin{tabular}{cc}
\includegraphics[scale=0.66]{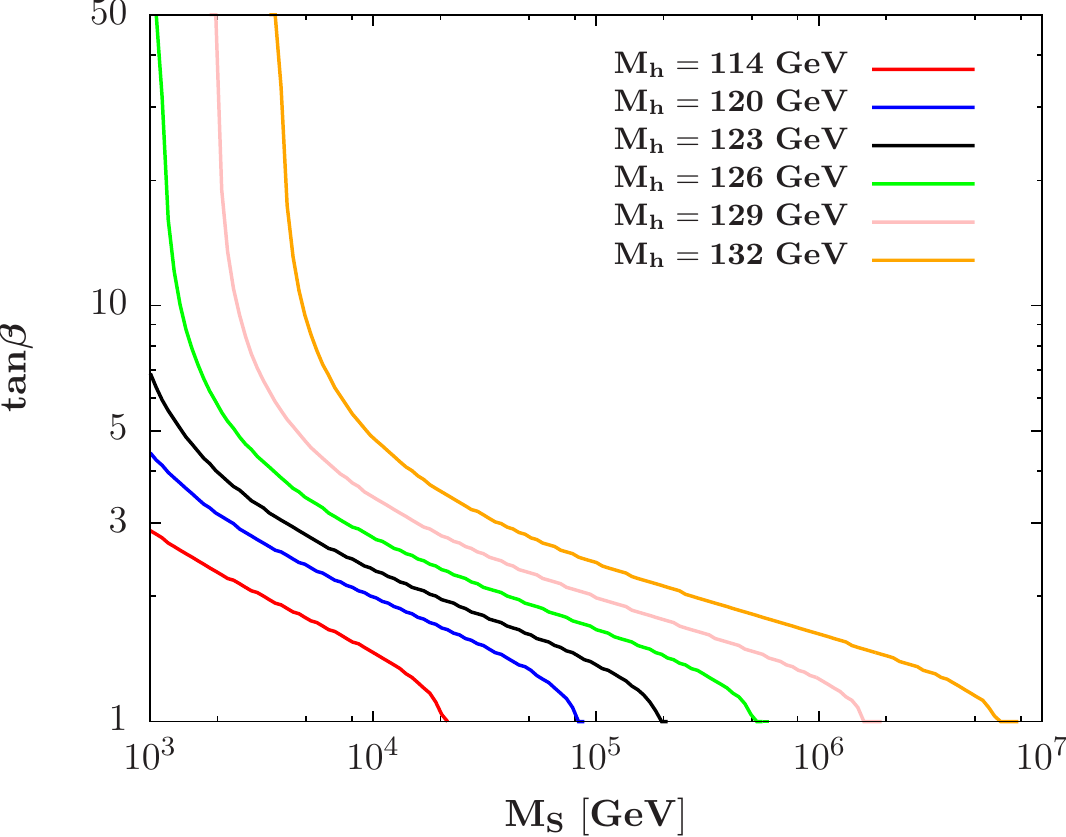} & \includegraphics[scale=0.66]{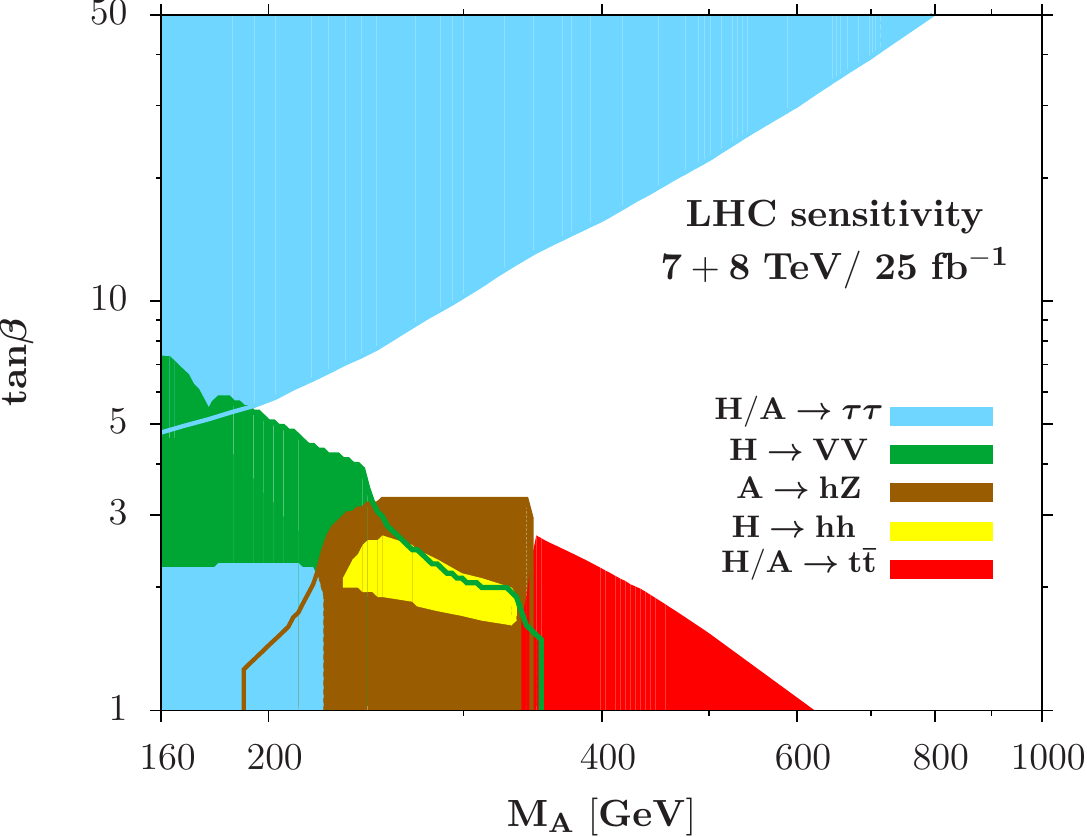}
\end{tabular}
\end{center}
%\vspace*{-7.2cm}
\caption[]{Left: contours for fixed values $M_h=120$--132 GeV in the $[\tb, M_{S}]$ 
plane in the decoupling limit $M_{A} \gg M_{Z}$; the ``LEP2 
contour" for $M_h=114$ GeV is shown in red. 
Right: the estimated sensitivities~\cite{O1} in the various search channels for the 
heavier MSSM Higgs bosons in the $[\tb,M_A]$ plane: $H/A \! \to \! \tau \tau$, 
$H \! \to \! WW\! +\! ZZ$, $H/A \! \to \! t\bar t$, $A\! \to \! hZ$  
and $H \! \to \! hh$. Taken from Ref.\cite{O1}.}
\label{fig2}
%\vspace*{-2mm}
\end{figure}

In our quite ``model--independent'' approach, defined in eq.~(\ref{wide}), we make no restriction on the SUSY scale which can be at any value, even quite high. It allows to reopen the small 
$\tan\beta$ region, $\tan\beta \! \lsim \! 3$, that was long thought to be excluded
from the negative search of a SM--like scalar boson at LEP which set the limit $M_{h} 
\! \gsim \! 114$ GeV, but assuming a setting with $M_S \! \lsim \! 1$~TeV.
If $M_{S}$ is large enough as indicated by present data (see Ref.\cite{Arbey:2011ab} for example), low $\tb$ values would still be allowed. 
In the left-hand side of Fig.~\ref{fig2}, we display the contours in the plane $[\tb, M_{S}]$ for mass values in the 
window $M_h=120$--132 GeV of the observed Higgs state. 

The contour corresponding to the LEP2 limit $M_h=114$ GeV indicates that 
 $\tb \approx 1$ is still viable provided that $M_S \gsim 20$~TeV. 
The present value $M_h=125$ sets stronger constraints: for example, while one can accommodate a scale $M_S \approx 1$ 
TeV with $\tb \approx 5$, a large scale $M_S \approx 20$~TeV is required to obtain $\tb \approx 2$. Let us discuss the implications for heavy Higgs searches. 

The most promising process to look for the heavier MSSM Higgs scalars is by far 
$pp \! \to \! gg\! +\! bb \! \to \! H/A\! \to \! \tau \tau$.  Searches for this 
channel have been performed by ATLAS~\cite{Aad:2012cfr} with $\approx 5$ fb$^{-1}$ data at the 7 TeV run and by CMS~\cite{CMS:2013hja} with $\approx 5+20$ fb$^{-1}$ data at the 7 TeV and
8 TeV runs. Upper limits on the production cross section times decay 
branching ratio have been set and they can be turned into constraints 
on the MSSM parameter space. 
The sensitivity of the CMS $pp\to h,H,A \to \tau \tau$ analysis in the plane $[\tb,M_A]$ using 25~fb$^{-1}$ of data can be found in Ref.\cite{CMS:2013hja}. The excluded region obtained from the observed limit at the 95\%CL is extremely restrictive and for $M_A \approx 250$~GeV the high $\tb \gsim 10$ region is entirely excluded and one is even sensitive to large values $M_A \approx 800$~GeV for  $\tb \gsim 45$. 

Nevertheless, there is a caveat to this exclusion limit because the constraint applies for a particular 
benchmark, the maximal mixing scenario with $X_t/M_S= \sqrt 6$, assuming $M_S=1$ TeV. 
In fact this exclusion limit is valid in far more situations than the ``MSSM $M_h^{\rm max}$ scenario" 
and it should be extended to the low $\tb$ regime which, 
in the chosen scenario with $M_{S}=1$ TeV, is excluded by the LEP2 limit on 
the lighter $h$ mass but is resurrected if the
SUSY scale is kept as a free parameter. 
Reopening the low $\tb$ region allows to hunt for the heavier scalar bosons in various interesting processes
at the LHC. Heavier CP--even $H$ decays into massive gauge  bosons $H\to WW,ZZ$ and 
lighter Higgs bosons $H\to hh$, CP--odd scalar decays into a vector and a Higgs boson, 
$A \to hZ$, CP--even and CP--odd scalar decays  into top quarks, $H/A \to t \bar 
t$, and the charged scalar decays into a gauge boson and a Higgs boson, $H^\pm \to Wh$. 

A preliminary study of these processes has been 
performed~\cite{O1} relying on the searches for the SM Higgs boson or other heavy resonances
made by the ATLAS and CMS collaborations.
The results which are shown on the left-hand of Fig.~\ref{fig2} are interesting 
since these searches cover a large part of the 
parameter space of the MSSM Higgs sector in a model--independent way, i.e. without the need to precise the 
SUSY particle spectrum that appear in the quantum corrections.
More especially, the channels $H \to VV$ and $H/A
\to t \bar t$ are very constraining as they probe the entire low
$\tb$ area up to $M_A \approx
600$ GeV. Notice that $A \to hZ$ and $H \to hh$ could also be seen at the current LHC 
in small parts of the MSSM parameter space.

\section{Summary}

We have discussed a simplified framework that describes the MSSM Higgs sector after
the discovery of the lighter $h$ boson. Including the constraint $M_h\!=\!125$ GeV, 
it can be again parameterized by the two inputs $\tb$ and $M_A$ as at 
tree-level, irrespective of the SUSY parameters that enter the radiative
corrections such as the SUSY scale $M_S$. Allowing large $M_S$ values reopens 
the low $\tb$ region which can be probed in many interesting processes at the LHC. 
This is the case of e.g. the processes $gg\to H/A \to t\bar t$ which need further
studies~\cite{Prepa1}.

\section*{Acknowledgments}\vspace*{-3mm}

This note relies on work performed in collaboration with A. Djouadi, L. Maiani, G. Moreau, A. Polosa and V. Riquer. 
The work has been done under the ERC Advance Grant Higgs@LHC and 
financial support from the Moriond organizing committee is acknowledged.

\end{document}